# Simulation of Stacks of High Temperature Superconducting Coated Conductors Magnetized by Pulsed Field Magnetization Using Controlled Magnetic Density Distribution Coils

Shengnan Zou, Victor Manuel Rodriguez Zermeno, Francesco Grilli

*Abstract*— High temperature superconducting (HTS) stacks of coated conductors (CCs) can work as strong trapped field magnets (TFMs) and show potential in electrical applications. Pulsed field magnetization (PFM) is a practical method to magnetize such TFMs, but due to heat generation during the dynamic process, it cannot achieve a trapped field as high as field cooling can. In this work, we construct a 2D electromagnetic-thermal coupled model to simulate stacks of HTS CCs with realistic laminated structures magnetized by PFM. The model considers temperature and anisotropic magnetic field dependent $J_c$ of HTS and other temperature dependent thermal and electrical material properties. Based on the model, a configuration of controlled magnetic density distribution coils is suggested to improve the trapped field compared to that obtained by ordinary solenoids.

## I. INTRODUCTION

HIGH temperature superconducting (HTS) bulks and stacks of coated conductors (CCs) can work as strong permanent magnets by trapping persistent currents [1-4]. To magnetize these trapped field magnets (TFMs), the pulsed field magnetization (PFM) technique is of special interest, because it can work *in situ* inside practical devices. However, the trapped field acquired by PFM is usually lower than that acquired by field cooling, because fast flux motions during the PFM generate considerable amount of heat and increase the temperature [5-7]. Studies by experiments and simulation have been reported to analyze the dynamics during the PFM and suggest strategies to improve the trapped field, but they mostly focus on HTS bulks [8-16]. In this study, a two-dimensional electromagnetic-thermal coupled model is constructed for stacks of HTS CCs magnetized by PFM. The model considers realistic laminated structures of coated conductors. For the first time, temperature dependence of anisotropic magnetic field dependent critical current density of HTS tapes and other material properties including heat capacity, thermal conductivity, and electrical resistivity are considered. With this model, we compare PFM using a solenoid with using controlled magnetic density distribution coils and show that the latter can improve the trapped field of a TFM.

## II. MODEL DESCRIPTION

The finite-element-method (FEM) model is a 2D model based on $H$-formulation of Maxwell equations and heat transfer equation implemented in COMSOL Multiphysics 5.0 [17]. The $H$-formulation of Maxwell equations has been widely utilized in simulating HTS materials [18-21]. It uses the General Form PDE module. The electrical property of HTS is described by the so-called $E$-$J$ power law [22]. In this work the exponent in the mentioned law is set to 21. The temperature and magnetic field dependence of critical current density $J_c$ will be described in Part III.

The thermal part of the model is implemented in the Heat Transfer in Solids module using the time-dependent heat transfer equation:

$$\rho \cdot C_p \frac{\partial T}{\partial t} = \nabla \cdot (\lambda \nabla T) + Q, \qquad (1)$$

where $\rho$ is the mass density, $C_p$ and $\lambda$ are temperature dependent heat capacity and thermal conductivity, respectively, and $Q$ equals $E \cdot J$, which is the heat generation power density (unit W/m$^3$) coupling the electromagnetic equation and the heat transfer equation. Langrange linear elements are used for discretization of the heat transfer equation to improve the computing speed. A Dirichlet boundary condition is used in the outer boundary of the stack to set a constant temperature of 30 K. A thermal resistive layer separates the boundary and the tape surface as indicated in Fig. 1. The layer is 1 mm thick with $\lambda$ equal 0.1 W/(m·K) in this model, which can be adjusted according to cooling efficiency [9-10].

In this work, we simulated stacks composed of 20-layer 12-mm wide tapes magnetized by PFM using two different coil configurations, which are a common solenoid and a set of controlled magnetic density distribution coils (CMDCs), as shown in Fig. 1. The latter is inspired by previous works on bulks which suggest that vortex coils have potential to improve the trapped field [10-11]. The CMDCs consist of a

This work has been partly supported by the Helmholtz Association (Young Investigator Group grant VH-NG-617).
S. Zou (e-mail: shengnan.zou@kit.edu), V. Zermeno (e-mail: victor.zermeno@kit.edu), and F. Grilli (e-mail: francesco.grilli@kit.edu), are with Institute for Technical Physics, Karlsruhe Institute of Technology, Hermann-von-Helmholtz-Platz 1, 76344 Eggenstein-Leopoldshafen, Germany.

pair of split coils. And each split coil consists of two coils carrying inverse currents. The inner coil with current $I_1$ generates the main applied field. The outer coil carries an inverse current $I_2=-\alpha I_1$ ($0<\alpha<1$) which weakens the applied field on the stack's periphery. The CMDCs thus can generate a highly non uniform field along the x axis, with a peak in the center. By adjusting $\alpha$, the gradient of the applied field can be adjusted, as shown in Fig. 2.

**FIG. 1 HERE**

**FIG. 2 HERE**

Considering symmetry, only one quarter of the geometry (the first quadrant in Fig. 1) is simulated in this work to save computing time.

The applied pulsed magnetic field is a time-dependent function given by,

$$B_{app} = B_0 \sin^2\left(\frac{\pi t}{2\tau}\right), \text{ when } t < \tau \quad (2a)$$

$$B_{app} = B_0 \cos^2\left(\frac{\pi(t-\tau)}{10\tau}\right), \text{ when } \tau \leq t \leq 6\tau. \quad (2b)$$

The applied field takes time $\tau$ to ramp to the peak value and uses $5\tau$ to damp to zero. We use $\tau$ equal 10 ms in this work. The function is arranged in such a way that the beginning and transition are smooth, which helps in numerical convergence. The shape of the pulse is shown in Fig. 3. The marked points will be used in part IV.

**FIG. 3 HERE**

### III. PARAMETERS

In this work, we consider comprehensive temperature dependent material properties for the first time including the critical density of HTS, electrical resistivity, heat capacity and thermal conductivity of all composing materials (except a 0.2 μm buffer layer). The composition of the tape is shown [25].

The critical current density ($J_c$) of HTS tapes depends on both the temperature and the local magnetic field. And its dependence on magnetic field is anisotropic. The elliptical equation is assumed [23],

$$L = \frac{J_c(\mathbf{B},T)}{J_c(77\text{ K})} = \frac{L_0(T)}{\left(1+\sqrt{k(T)^2 B_x^2 + B_y^2}/B_{c0}(T)\right)^{b(T)}}, \quad (3)$$

where $L$ is the lift factor, representing the critical current density divided by its self-field value at 77 K. The parameters $k(T)$, $B_{c0}(T)$, $b(T)$ are temperature dependent to reflect a varying in-field property of $J_c$ with temperature.

To make reasonable assumptions of $J_c$ in this work, we take lift factor measurement data of YBCO superconducting tapes from SuperPower Inc. [24]. The critical current density is obtained by dividing the measured critical current by the cross-sectional area of the HTS layer. This is considered to be a valid approach, because the PFM works in strong magnetic field and the self-field influence can be neglected. Parameters for each temperature are fitted with Eq. (3) with the least root mean square error value. The results for each temperature are shown in Tab. 1. Each parameter varies smoothly with temperature, so parameters for arbitrary temperature are estimated by directly interpolating between temperatures. In this work, $J_c$(sf, 77 K) equals 3 MA/cm$^2$.

**TAB. 1 HERE**

Other temperature dependent material properties, including electrical resistivity ($\rho_e$), heat capacity ($C_p$) and thermal conductivity ($\lambda$) are also considered in this model by directly using interpolated data from experiments [26-31]. The values are shown in Fig. 4. The residual resistance ratio (RRR) equal 42 is used as suggested in [27] for copper, which is the dominant material in the tapes. The temperature dependence of $C_p$ is especially important because it avoids unrealistic over-rising of the temperature and improves the numerical stability of the model.

**FIG. 4 HERE**

### IV. RESULTS AND DISCUSSION

The stacks are magnetized by pulsed field with various magnitudes using one infinite solenoid and three different CMDCs as described in Fig. 1 and 2. The trapped field is measured in the center of the sample (x=0 mm) and 0.8 mm above the surface. It is evaluated 10 s after the pulse, which allows the stack to relax fully and cool down to 30 K. The results are shown in Fig. 5.

**FIG. 5 HERE**

For each coil, the trapped field first increases and then decreases with the applied field. This is consistent with numerous previous theoretical and experimental works on bulks [7-14]. Too low applied field cannot induce full currents in the sample; however, too high applied field will generate excessive heat, increase the temperature and reduce $J_c$. There is an optimal applied field, which results in the maximum trapped field. Interestingly, the maximum trapped field acquired by CMDCs is larger than that by the solenoid. And the larger the gradient of the applied field generated by CMDCs, the larger the maximum trapped field. This finding shows consistency with [10-11], which finds that split coils may generate larger trapped field in HTS bulks compared to solenoids.

**FIG. 6 HERE**

The penetration processes during PFM by solenoid and CMDC-medium are compared in Fig. 6. The normalized critical current density $J/J_c$ is plotted. The applied fields in this plot are of magnitudes (2.6 T for the solenoid and 3.2 T for the CMDC-medium) when the maximum trapped fields are obtained (marked points in Fig.5). The selected time points were illustrated in Fig. 3. As shown in (f), the maximum trapped field is acquired when the stack is not fully filled with positive currents. This suggests that for PFM at low temperature, to use the full cross-section of the sample to carry remanent currents will generate excessive heat and have negative influence. In (f) $J/J_c$ is only around 0.5, suggesting a loss of trapped field due to dynamics during PFM compared to field cooling. From (e) to (f), currents first decay quickly and then stabilize ($J$ data not shown). And $J_c$ increases when the temperature is decreased to 30 K.

Comparing the two coils, the penetration of CMDCs starts from the surface of the stack and the current front tends to parallel to the surface, unlike that of the solenoid. This will result in less heat generation on the periphery of the stack compared to the solenoid case. The magnetic field and flux lines distribution at time (a) are shown in Fig. 7. For the solenoid, the flux lines tend to accumulate on the periphery of the stack; for the CMDCs, the flux lines arrange parallel to the stack surface and the magnitude of the magnetic fields inside the stack is lower. Moreover, the field in the CMDCs situation is mainly in the direction perpendicular to the c-axis of the tape. In this way, the $J_c$ is less degraded according to Eq. (3) in the CMDCs case. This also partially explains the reason why CMDCs can increase the trapped field.

**FIG. 7 HERE**

The heat generation powers of different composite components during the PFM are plotted in Fig. 8. The total heat generation (unit J/m in 2D) during PFM is given in Tab.2, which is the integration of the curves in Fig. 8. Generally, the heat generation powers are larger in the pulse ascending stage compared to the descending stage, because the former has a larger field changing rate. For both coils, the superconducting layers generate the most heat. Copper layers also generate substantial heat due to eddy currents. Silver layers generate less heat because of their small thickness. The heat generation of Hastelloy is considerably lower thanks to its large electrical resistivity. As a result, tapes with metallic stabilizers of larger resistivity may be more suitable for application in TFMs because they generate less heat during PFM.

**FIG. 8 HERE**

**TAB. 2 HERE**

Comparing Solenoid and CMDC-medium, the heat generation power is less for CMDC-medium, not only for the superconducting components, but also for the metals. The peaks of heat generation power do not happen at the same time, because of their different penetration processes. At time (e), the maximum temperature is 50 K for CMDC-medium, while 56 K for solenoid. In this way, CMDCs reduce the heat generation and lower the temperature rise compared to solenoid, so that the final trapped field can be increased.

## V. CONCLUSION

In this work, we construct a 2D electromagnetic-thermal coupled model for stacks of HTS coated conductors magnetized by PFM. The model considers the realistic geometry of all components of the HTS tapes, temperature and anisotropic magnetic field dependent critical current density of HTS and temperature dependent thermal and electrical parameters of all composite materials. Based on the model, PFM by solenoid and controlled magnetic density distribution coils are compared. The main findings are: a) temperature and anisotropic magnetic field dependent critical current density of HTS and temperature-dependent material parameters, especially heat capacity should be considered in numerical models, because they change dramatically with temperature during the PFM; b) less conductive stabilizers, for example coppers of smaller RRR, are more favorable for stacks magnetized by PFM, because they generate less heat; c) CMDCs, which generate gradient applied fields with peak values in the center, can reduce heat generation during PFM and increase the trapped field; the larger the gradient, the larger the increase (within the extent of this work). This work proposes the possibility of increasing the trapped field by PFM using different coil configurations; however, it requires more complex engineering to build such coils. Further work needs to be done for evaluating the applicability of this strategy.

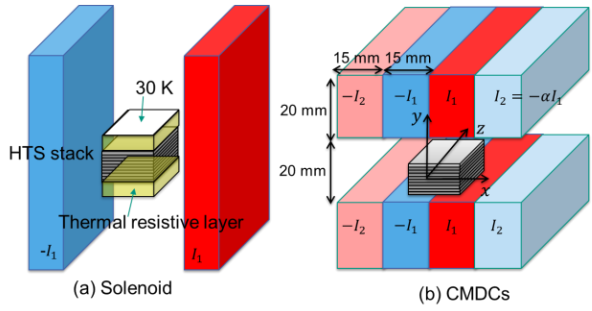

Fig. 1. Schematics of the coils used for the PFM of HTS stacks. (a) Solenoid; (b) Controlled magnetic density distribution coils (CMDCs). The thermal resistive layers and coordinates apply for both (a) and (b).

TABLE I
FITTED PARAMETERS FOR THE LIFT FACTOR EQUATION

| $T$ (K) | $L_0$ | k | $B_{c0}$ (T) | b |
|---|---|---|---|---|
| 20 | 6.52 | 0.06 | 5.04 | 1.64 |
| 30 | 6.12 | 0.07 | 3.23 | 1.41 |
| 40 | 5.29 | 0.10 | 1.86 | 1.12 |
| 50 | 4.12 | 0.17 | 1.26 | 0.96 |
| 65 | 2.44 | 0.61 | 0.59 | 0.77 |

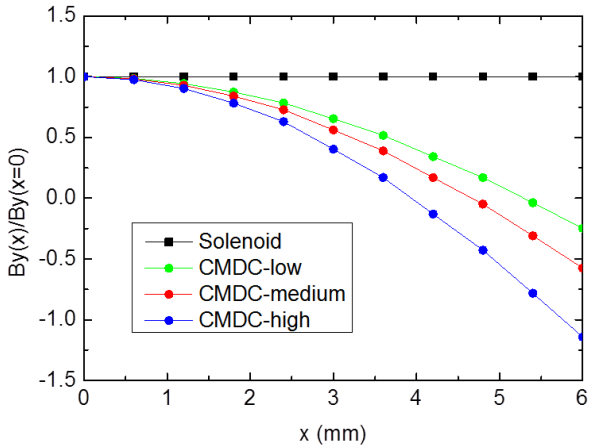

Fig. 2. The magnetic field density distribution of applied field of coils along the tape width (from x=0 to 6 mm) in Fig.1. The solenoid produces a uniform field. CMDCs produce fields with different gradients. CMDC-low, CMDC-medium and CMDC-high correspond to α equal 0.6, 0.62 and 0.64 respectively.

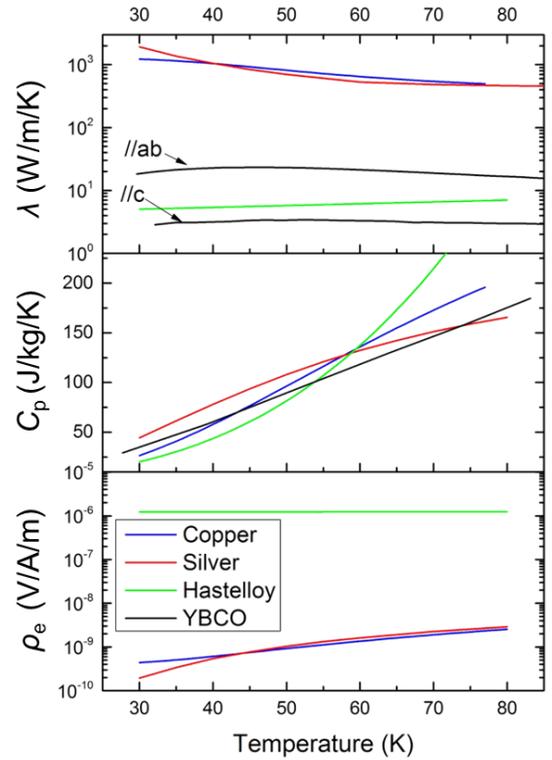

Fig. 4. Temperature dependent thermal conductivity (λ), heat capacity ($C_p$) and electrical resistivity ($\rho_e$) of composing materials of HTS tapes.

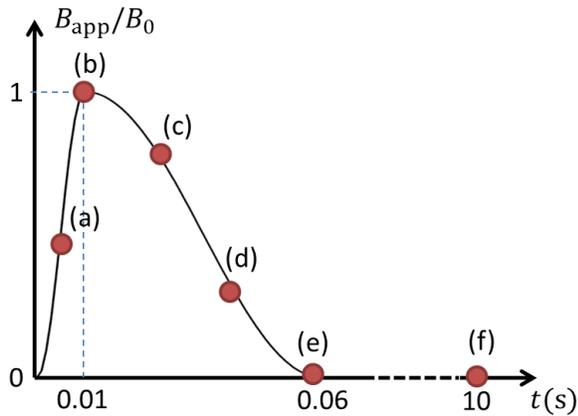

Fig. 3. The shape of applied pulse as described in Eq. (2). Selected points for analysis are: (a) 0.005 s (b) 0.01 s (c) 0.015 s (d) 0.04 s (e) 0.06 s (f) 10 s

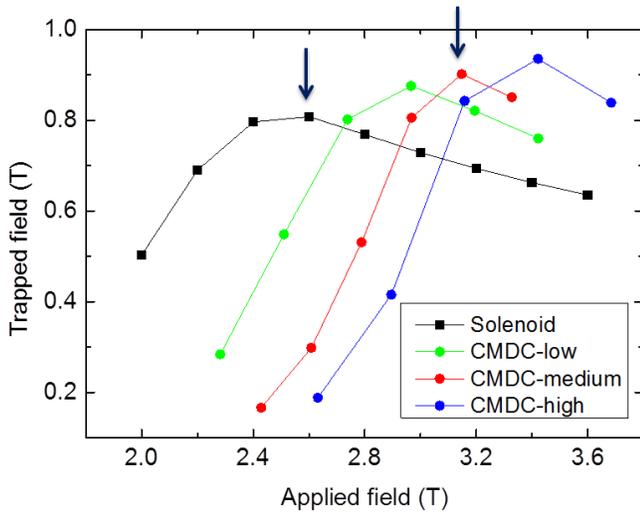

Fig. 5. The trapped field measured 0.8 mm above the center of the sample 10 s after the pulse with different magnitudes. The four lines show the results using different coils as described in Fig. 1 and 2.

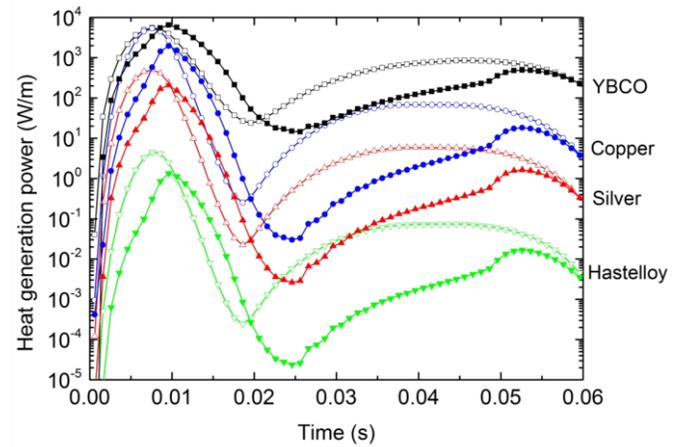

Fig. 8. Heat generation power of different components during PFM. Empty dots and solid dots represent Solenoid and CMDC-medium, respectively. Four different colours represent four different composite materials.

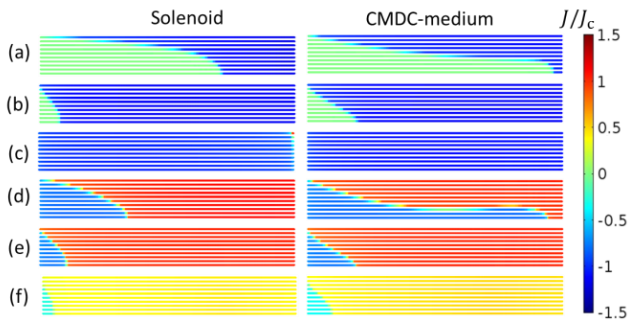

Fig. 6. Comparison of penetration processes during PFM between solenoid and CMDC-medium. The colour scale shows the normalized current density $J/J_c$. (a) to (f) correspond to selected points as illustrated in Fig. 5. The plots show one quarter of the sample. The thickness of each tape is exaggerated for better visualization.

TABLE II
TOTAL HEAT GENERATION OF DIFFERENT COMPOSITE MATERIALS DURING PFM

| Total heat generation (J/m) | Solenoid | CMDC-medium |
| --- | --- | --- |
| YBCO | 52.0 | 44.8 |
| Copper | 24.1 | 7.9 |
| Silver | 2.4 | 0.9 |
| Hastelloy | 0.0 | 0.0 |

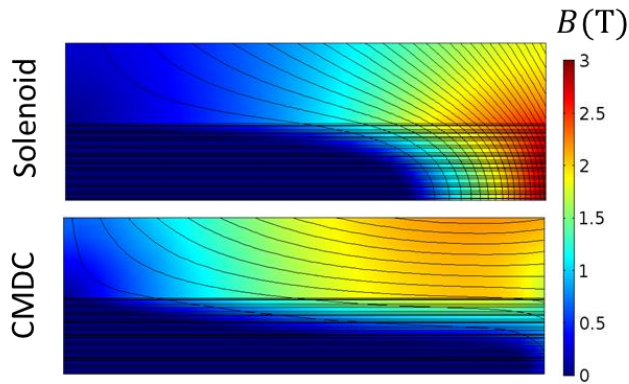

Fig. 7. Comparison of the magnetic field distribution and flux lines in the stack magnetized by the solenoid and CMDC-medium at time (a). The plots show one quarter of the sample.